# Learning Semantic String Transformations from Examples

Rishabh Singh[*]
MIT CSAIL, Cambridge, MA
rishabh@csail.mit.edu

Sumit Gulwani
Microsoft Research, Redmond, WA
sumitg@microsoft.com

## ABSTRACT

We address the problem of performing semantic transformations on strings, which may represent a variety of data types (or their combination) such as a column in a relational table, time, date, currency, etc. Unlike syntactic transformations, which are based on regular expressions and which interpret a string as a sequence of characters, semantic transformations additionally require exploiting the semantics of the data type represented by the string, which may be encoded as a database of relational tables. Manually performing such transformations on a large collection of strings is error prone and cumbersome, while programmatic solutions are beyond the skill-set of end-users. We present a *programming by example* technology that allows end-users to automate such repetitive tasks.

We describe an expressive transformation language for semantic manipulation that combines table lookup operations and syntactic manipulations. We then present a synthesis algorithm that can learn all transformations in the language that are consistent with the user-provided set of input-output examples. We have implemented this technology as an add-in for the Microsoft Excel Spreadsheet system and have evaluated it successfully over several benchmarks picked from various Excel help-forums.

## 1. INTRODUCTION

The IT revolution over the past few decades has resulted in two significant advances: digitization of massive amounts of data and accessibility of computational devices to massive proportions of the population. It is thus not surprising that more than 500 million people worldwide use spreadsheets for storing and manipulating data. These business *end-users* have myriad diverse backgrounds and include commodity traders, graphic designers, chemists, human resource managers, finance pros, marketing managers, underwriters, compliance officers, and even mail room clerks—they are not professional programmers, but they need to create small, *often one-off*, applications to support business functions [6].

Unfortunately, the programming experience since inception has mostly focused on serving the needs of a select class of few million skilled users. In particular, spreadsheet systems like Microsoft Excel allow sophisticated users to write macros using a rich inbuilt library of string and numerical functions, or to write arbitrary scripts using a variety of programming languages like Visual Basic, or .NET. Since end-users are not proficient in programming, they find it too difficult to write desired macros or scripts.

The combination of the above-mentioned technical trends and lack of a satisfactory solution has led to a marketplace of hundreds of advertisement-driven help-forums[1], some of which contain millions of posts from end-users soliciting help for scripts to manipulate data in their spreadsheets. The experts respond to these requests after some time. From an extensive case study of such spreadsheet help-forums, we observed the following two things.

**Semantic String Transformations:** Several of the requested scripts/macros were for manipulating strings that need to be interpreted as more than a sequence of characters, e.g., as a column entry from some relational table, or as some standard data type such as date, time, or currency. (See §2 for a motivating example.) In this paper, we describe the systematic design of a semantic transformation language for manipulating such strings.

**Input-Output Examples based Interaction Model:** End-users used input-output examples as the most common and natural way of expressing intent to experts on the other side of the help-forums. An expert provides a program/transformation that is consistent with those examples. If the end-user is not happy with the result of the program on any other new input in the spreadsheet, the interaction is repeated with an extended set of input-output examples. This is the natural interface that our tool provides to the end-user. We describe the systematic design of an (inductive) synthesis algorithm that can learn desired scripts in our transformation language from very few examples.

We observe that most *semantic transformations* can be expressed as a combination of *lookup transformations* and *syntactic transformations*. We use this observation to present a systematic design of the transformation language for performing semantic string transformations. We first present an expressive language for lookup transformations and then extend it by adding syntactic transformations [8].

---

[*]Work done during two internships at Microsoft Research.

[1]http://www.excelforum.com/, http://www.ozgrid.com/forum/, http://www.mrexcel.com/, http://www.exceltip.com/

| Input $v_1$ | Input $v_2$ | Output |
|---|---|---|
| Stroller | 10/12/2010 | $145.67+0.30*145.67 |
| Bib | 23/12/2010 | $3.56+0.45*3.56 |
| Diapers | 21/1/2011 | **$21.45+0.35*21.45** |
| Wipes | 2/4/2009 | **$5.12+0.40*5.12** |
| Aspirator | 23/2/2010 | **$2.56+0.30*2.56** |

| MarkupRec | | |
|---|---|---|
| Id | Name | Markup |
| S30 | Stroller | 30% |
| B56 | Bib | 45% |
| D32 | Diapers | 35% |
| W98 | Wipes | 40% |
| A46 | Aspirator | 30% |
| ... | ... | ... |

| CostRec | | |
|---|---|---|
| Id | Date | Price |
| S30 | 12/2010 | $145.67 |
| S30 | 11/2010 | $142.38 |
| B56 | 12/2010 | $3.56 |
| D32 | 1/2011 | $21.45 |
| W98 | 4/2009 | $5.12 |
| A46 | 2/2010 | $2.56 |
| ... | ... | ... |

**Figure 1: A transformation that requires perfoming syntactic manipulations on multiple lookup results.**

We also describe a systematic design of the synthesis algorithm for the semantic string transformation language, which can synthesize a set of semantic transformations that are consistent with the given set of input-output examples. We first describe a synthesis algorithm for the lookup transformation language $L_t$, and then extend it to a synthesis algorithm for the extension of $L_t$ with syntactic transformations. Experimental results on our benchmark examples show that our algorithm is scalable and can learn desired transformations from very few examples.

This paper makes the following key contributions:

- We describe a lookup transformation language $L_t$ and an inductive synthesis algorithm for it (§4).
- We observe that most semantic transformations can be expressed as combination of lookup and syntactic transformations. We extend language $L_t$ with syntactic transformations to obtain a very expressive language $L_u$ and describe an inductive synthesis algorithm for it (§5).
- We describe an experimental prototype of our system that has an attractive user-interface and is ready to be deployed. We present the evaluation of our prototype on a large collection of benchmarks obtained from help-forums, books, mailing lists and Excel product team (§7).

This paper is organized as follows. §3 establishes a common framework for describing transformation languages and their inductive synthesis algorithms. §4 and §5 describe novel instantiations of this framework for a lookup transformation language and for its extension with a syntactic transformation language. §6 shows applications of the semantic transformation language (described in §5) for performing transformations on standard data types.

## 2. MOTIVATING EXAMPLE

Consider the following example from an Excel help-forum.

EXAMPLE 1. *A shopkeeper wanted to compute the* **selling price** *of an item (Output) from its* **name** *(Input $v_1$) and* **selling date** *(Input $v_2$) using the* **MarkupRec** *and* **CostRec** *tables as shown in Figure 1. The selling price of an item is computed by adding its purchase price (for the corresponding month) to its markup charges, which in turn is calculated by multiplying the markup percentage by the purchase price.*

The user expresses her intent by giving a couple of examples (i.e., the first two rows). Our tool then automates the repetitive task (i.e., fills in the bold entries). We highlight below the key technical challenges involved.

*Expressive transformation language.* The transformation/program inferred by our system for automating the repetitive task involves both lookup and syntactic operations. In particular, note that we need lookup operations for (i) obtaining the markup percentage from an item name in MarkupRec table (Stroller → 30%) and (ii) for obtaining the purchase price of item in CostRec table after performing a join operation between the two tables on the item Id column (Stroller,12/2010 → $145.67). Observe that the string 12/2010 used for performing the second lookup is obtained by performing a syntactic transformation (namely, a substring operation) on the input string 10/12/2010. After performing the lookups, we need a syntactic transformation (namely, a concatenate operation) to concatenate the lookup outputs with constant strings like +,0.,* in a particular order to generate the final output string. We present an expressive transformation language that combines lookup and syntactic transformations in a nested manner.

*Succinct representation & efficient computation of large number of consistent transformations.* The number of expressions in our expressive transformation language that are consistent with a given input-output example can potentially be very large. For example, for the first input-output example (Stroller, 10/12/2012 → $145.67+0.30*145.67), there are a large number of transformations that can generate the output string. In general, every substring in the output string can potentially be either a constant string, or a substring of an input string, or the result of a lookup operation. For example, the substring 30 in the output string can either be a constant string or a string obtained by performing a lookup operation in the MarkupRec or CostRec table. Some of the possible lookup transformations to obtain the string 30 include selecting the Markup column entry in the MarkupRec table where the item Name in the corresponding row is either one of the constant strings Stroller or Aspirator, or it matches the input string $v_1$. Another valid lookup transformation is to select the last two characters from the item Id column (S30) in MarkupRec or CostRec table with various ways to select the first row by constraining the item Name or Date columns respectively. We thus have a large number of possible transformations for each substring of the output string and explicit enumeration of all such choices becomes infeasible. A key technical contribution of this paper is a data structure that can succinctly represent an exponential number of such transformations in polynomial space, and an algorithm that can compute such transformations in polynomial time. The key idea is to share common sub-expressions and compute/maintain choices for independent sub-expressions independently.

*Ranking.* Our synthesis algorithm learns the set of all consistent transformations for each example and then intersects these sets to obtain the common transformations. The number of examples required to converge to the desired transformation may be large. To enable learning of the desired transformation from very few examples, we perform a ranking of these learned transformations that gives preference to transformations that are smaller (Occam's razor principle) and that use fewer constants (to enforce generalization).

## 3. OVERVIEW

We describe our expressive transformation language and associated data structures and algorithms for inductive synthesis in two steps (§4 and §5). In this section, we introduce the common formalism and our user interaction model (as derived from recent work on inductive synthesis [8, 9]).

### 3.1 Formalism

*Transformation language $L$.* The first step in inductive synthesis is to define a domain specific language that is expressive enough to capture several real-world tasks, but at the same time is restrictive enough to enable efficient learning from input-output examples. In this paper, we introduce a string transformation expression language $L$, which contains expressions $e$ that map an input state $\sigma$, which holds values for $m$ string variables $v_1, .., v_m$ (denoting the multiple input columns in a spreadsheet), to an output string $s$.

$$e : (\texttt{String} \times \ldots \times \texttt{String}) \to \texttt{String}$$

This formalism can also be used for string processing tasks that require generating a tuple of $n$ strings as output by solving $n$ independent problems. We characterize an expression language $L$ with the following components: (a) a set of grammar rules $R$, (b) a top-level symbol $e$, which is a uniquely distinguished non-terminal symbol occurring in $R$. §4.1 and §5.1 describe two examples of such a language.

*Data structure $D$ for set of expressions in $L$.* The number of transformation expressions that are consistent with a given set of input-output examples can be huge. We define a data structure $D$ to succinctly represent such a large set of expressions. We describe $D$ itself using a set of grammar rules $\tilde{R}$ with top-level symbol $\tilde{e}$. §4.2 and §5.2 describe examples of such a data structure that uses a top-level graph/DAG representation respectively.

*Synthesis algorithm:* `GenerateStr` *and* `Intersect`. The inductive synthesis algorithm `Synthesize` for an expression language $L$ learns the set of expressions in $L$ (represented using data structure $D$) that are consistent with a given set of input-output examples. Our synthesis algorithm consists of the following two procedures:

- The `GenerateStr` procedure for computing the set of all expressions (represented using data structure $D$) that are consistent with a given input-output example.
- The `Intersect` procedure for intersecting two sets of expressions (represented using data structure $D$). We describe `Intersect` procedure also using a set of rules.

DEFINITION 1. *(Soundness/k-completeness of* `GenerateStr`*)* Let $\tilde{e} = $ `GenerateStr`$(\sigma, s)$. *We say that* `GenerateStr` *procedure is* sound *if all expressions in $\tilde{e}$ are consistent with the input-output example $(\sigma, s)$. We say that* `GenerateStr` *procedure is* k-complete *if $\tilde{e}$ includes all expressions with at most k recursive sub-expressions that are consistent with the input-output example $(\sigma, s)$.*

DEFINITION 2. *(Soundness/Completeness of* `Intersect`*)* Let $\tilde{e}'' = $ `Intersect`$(\tilde{e}, \tilde{e}')$. *We say that* `Intersect` *is* sound and complete *iff $\tilde{e}''$ includes all expressions that belong to both $\tilde{e}$ and $\tilde{e}'$.*

§4.3 and §5.3 give examples of `GenerateStr`/`Intersect` procedures that are sound and k-complete/complete and that have polynomial time complexity.

The synthesis algorithm `Synthesize` involves invoking the `GenerateStr` procedure on each input-output example, and intersecting the results using the `Intersect` procedure:

$\underline{\texttt{Synthesize}((\sigma_1, s_1), \ldots, (\sigma_n, s_n))}$
1   $P := \texttt{GenerateStr}(\sigma_1, s_1);$
2   for $i$ = 2 to $n$:
3     $P' := \texttt{GenerateStr}(\sigma_i, s_i); P := \texttt{Intersect}(P, P');$
4   return $P$;

*Ranking.* An expressive domain-specific language $L$ for inductive synthesis can often require a large number of examples to learn the intended transformation. We address this problem by developing a ranking scheme that can be used to rank the possibly large number of transformation expressions that are consistent with a small number of input-output examples. This ranking scheme is inspired by the Occam's razor principle, which states that a smaller and simpler explanation is usually the correct one. We define a comparison scheme between different expressions by defining a partial order between them. Any partial order can be used that is consistent with the sharing used in data structure $D$ for succinct representation of sets of such expressions. In other words, the comparison of any two sub-expressions should be based only on attributes that are not shared at the level of the sub-expressions. This allows us to efficiently identify the top ranked expressions from among a set of expressions represented using $D$. We give examples of such a ranking scheme in §4.4 and §5.4. Some of these choices are subjective, but our experiments illustrate that our ranking scheme works very effectively in practice: all of our benchmark tasks required at most 3 input-output examples.

### 3.2 User Interaction Model

The user expresses her intent of the task using few input-output examples. The synthesizer based on the above formalism then generates a ranked set of transformations that are consistent with those examples. We describe below some interaction techniques for automating the desired task or for generating a reusable transformation.

The user can run the top-ranked synthesized transformation on other inputs in the spreadsheet and check the generated outputs. If any output is incorrect, the user can fix it and the synthesizer can repeat the learning process with the additional example that the user provided as a fix. Requiring the user to check the results of the synthesizer, especially on a large spreadsheet, can be cumbersome. To enable easier interaction, the synthesizer can run *all* transformations on each new input to generate a set of corresponding outputs for that input. The synthesizer can then highlight those inputs (for user inspection) whose corresponding output set contains at least two outputs. The user can then focus their inspection on the highlighted inputs. Our prototype, implemented as an Excel add-in, supports this interaction model (which is also used in [8]).

On the other hand, if the user wishes to learn a reusable script, then the synthesizer may present the set of synthesized transformations to the user. Either the top-k transformations can be shown, or the synthesizer can walk the user through the data structure that succinctly represents all consistent transformations and let the user select the desired one. The transformations can be shown using the surface syntax, or can be paraphrased in a natural language. The differences between different transformations can also be explained by synthesizing a *distinguishing input* on which the transformations behave differently [11]. After receiving the correct output from the user on the distinguishing input, the synthesizer can repeat the learning process with this additional example.

## 4. LOOKUP TRANSFORMATIONS

In this section, we present a lookup transformation language $L_t$ that can model transformations that involve mapping a tuple of strings to another string using (possibly nested) lookup operations over a given database of relational tables. We first present the syntax and semantics of $L_t$ and then present a data structure $D_t$ to succinctly represent a large set of expressions in the language. We then present an efficient synthesis algorithm to learn a set of transformations in $L_t$ from a set of input-output examples, such that each of the learned transformations when run on the given inputs produces the corresponding outputs.

### 4.1 Lookup Transformation Language $L_t$

The syntax of our expression language $L_t$ for lookup transformations is defined in Figure 3(a). An expression $e_t$ is either an input string variable $v_i$, or a select expression denoted by $\text{Select}(C, T, b)$, where $T$ is a relational table identifier and $C$ is a column identifier of the table. The Boolean condition $b$ is an ordered conjunction of predicates $p_1 \wedge \ldots \wedge p_n$ where predicate $p$ is an equality comparison between the content of some column of the table with a constant or an expression. We place a restriction on the columns present in these conditional predicates namely that these columns together constitute a *candidate key* of the table. The main idea behind this restriction is that we want to express queries that produce a single output as opposed to a set of outputs. The ordering of predicates results in an efficient *intersection* algorithm as described in §4.3.

The language $L_t$ has expected semantics. The expression $\text{Select}(C, T, b)$ returns the table entry $T[C, r]$, where $r$ is the only row that satisfies condition $b$ (as condition $b$ is over candidate keys of the table). If there exists no row $r$ whose columns satisfy $b$, the expression returns the empty string $\epsilon$. The predicate $C = e_t$ is evaluated for row $r$ by first evaluating the expression $e_t$ and then comparing the returned string $[\![e_t]\!]\sigma$ with the string $T[C, r]$.

We now present an example taken from an Excel help-forum that can be represented in $L_t$.

EXAMPLE 2. *An Excel user was working on two tables: `CustData` and `Sale`. The user wanted to map names of customers to the selling price using address and street number columns as common columns between the two tables and posted the example shown in Figure 2 on a help-forum.*

The transformation can be expressed in $L_t$ as
$\text{Select}(\text{Price}, \text{Sale}, \text{Addr} = \text{Select}(\text{Addr}, \text{CustData},$
  $\text{Name} = v_1) \wedge \text{St} = \text{Select}(\text{St}, \text{CustData}, \text{Name} = v_1))$.

The surface syntax of $L_t$ allows sharing of sub-expressions (which is the key principle used in data structure $D_t$ described in §4.2). To appreciate this, consider the following example, which is also our running example in this section.

EXAMPLE 3. *Consider $m$ tables $T_1$ to $T_m$, each containing three columns $C_1$, $C_2$, and $C_3$ with $C_1$ being the primary key. Suppose table $T_i$ contains a row $(s_i, s_{i+1}, s_{i+2})$. Now given an input-output example $s_1 \to s_m$, we want to compute all expressions in $L_t$ that are consistent with it.*

| Input $v_1$ | Output |
|---|---|
| Peter Shaw | 110 |
| Gary Lamb | **225** |
| Mike Henry | **2015** |
| Sean Riley | **495** |

| CustData | | | | Sale | | | |
|---|---|---|---|---|---|---|---|
| Name | Addr | St | | Addr | St | Date | Price |
| Sean Riley | 432 | 15th | | 24 | 18th | 5/21 | 110 |
| Peter Shaw | 24 | 18th | | 104 | 12th | 5/23 | 225 |
| Mike Henry | 432 | 18th | | 432 | 18th | 5/20 | 2015 |
| Gary Lamb | 104 | 12th | | 432 | 15th | 5/24 | 495 |
| ... | ... | ... | | ... | ... | ... | ... |

**Figure 2: A lookup transformation that requires joining two tables.**

Consider the case of $m = 4$. Let $\mathbf{e} \equiv \text{Select}(C_2, T_1, C_1 = v_1)$ that produces string $s_2$. Two possible expressions in $L_t$ to obtain output $s_4$ from input $s_1$ are: (i) $e_1 \equiv \text{Select}(C_3, T_2, C_1 = \mathbf{e})$ (corresponding to path $s_1 \to s_2 \to s_4$) and (ii) $e_2 \equiv \text{Select}(C_2, T_3, C_1 = \text{Select}(C_2, T_2, C_1 = \mathbf{e}))$ (corresponding to path $s_1 \to s_2 \to s_3 \to s_4$). The expressions $e_1$ and $e_2$ share the common sub-expression $e$ which corresponds to obtaining the intermediate string $s_2$.

### 4.2 Data Structure for Set of Expressions in $L_t$

The set of expressions in language $L_t$ that are consistent with a given input-output example can be exponential in the number of *reachable* table entries. We represent this set succinctly using the data structure $D_t$, which is described in Figure 3(b). The data structure consists of a generalized expression $\tilde{e}_t$, generalized Boolean condition $\tilde{b}$, and generalized predicate $\tilde{p}$ (which respectively denote a set of $L_t$ expressions, a set of Boolean conditions $b$, and a set of predicates $p$). The formal semantics $[\![.]\!]$ of the data structure is described in Figure 3(c). The generalized expression $\tilde{e}_t$ is represented using a tuple $(\tilde{\eta}, \eta^t, \text{Progs})$ where $\tilde{\eta}$ is a set of nodes containing a distinct target node $\eta^t$ (representing the output string), and $\text{Progs} : \tilde{\eta} \to 2^{\tilde{f}}$ maps each node $\eta \in \tilde{\eta}$ to a set consisting of input variables $v_i$ or generalized select expressions $\text{Select}(C, T, B)$. A generalized select expression takes a set of generalized Boolean conditions $\tilde{b}_i$ as the last argument, where each $\tilde{b}_i$ corresponds to some candidate key of table $T$. A generalized conditional $\tilde{b}$ is a conjunction of generalized predicates $\tilde{p}_i$, where each $\tilde{p}_i$ is an equality comparison of the $j^{th}$ column of the corresponding candidate key with a constant string $s$ or some node $\tilde{\eta}$ or both. There are two key aspects of this data structure which are explained below using some worst-case examples.

**Use of intermediate nodes in $\tilde{\eta}$ for sharing:** Consider the problem in Example 3. The set of all transformations in $L_t$ that are consistent with the example $s_1 \to s_m$ can be represented succinctly using our data structure as: $\{\{\eta_1, \ldots, \eta_m\}, \eta_m, \text{Progs}\}$, where $\text{Progs}[\eta_i] = \{\text{Select}(C_2, T_{i-1}, \{C_1 = \{s_{i-1}, \eta_{i-1}\}\}), \text{Select}(C_3, T_{i-2}, \{C_1 = \{s_{i-2}, \eta_{i-2}\}\})\}$, $\text{Progs}[\eta_1] = \{v_1\}$, and $\text{Progs}[\eta_2] = \{\text{Select}(C_2, T_1, \{C_1 = \{s_1, \eta_1\}\})\}$. The node $\eta_i$ essentially corresponds to the string $s_i$. Figure 4 illustrates how the various nodes can be reached or computed from one another. Let $\text{N(i)}$ denote the number of expressions represented succinctly by $\text{Progs}[\eta_i]$. We have $\text{N(i)} = 2 + \text{N(i-1)} + \text{N(i-2)}$, implying that $\text{N(i)} = \Theta(2^i)$. Observe how our data structure

743

$$
\begin{array}{rcl}
\text{Expression } e_t & := & v_i \\
& | & \texttt{Select}(C,T,b) \\
\text{Boolean Cond } b & := & p_1 \wedge \ldots \wedge p_n \\
\text{Predicate } p & := & C = s \\
& | & C = e_t
\end{array}
$$
(a)

$$
\begin{array}{rcl}
\tilde{e}_t & := & (\tilde{\eta}, \eta^t, \texttt{Progs}) \\
& & \text{where } \texttt{Progs} : \tilde{\eta} \to 2^{\tilde{f}} \\
\tilde{f} & := & v_i \mid \texttt{Select}(C,T,B) \\
B & := & \{\tilde{b}_i\}_i \\
\tilde{b} & := & \tilde{p}_1 \wedge \ldots \wedge \tilde{p}_n \\
\tilde{p} & := & C = s \mid C = \eta \\
& | & C = \{s, \eta\}
\end{array}
$$
(b)

$$
\begin{array}{rcl}
[\![(\tilde{\eta}, \eta^t, \texttt{Progs})]\!] & = & \{e_t \mid e_t \in [\![\tilde{f}]\!], \tilde{f} \in \texttt{Progs}[\eta^t]\} \\
[\![v_i]\!] & = & \{v_i\} \\
[\![\texttt{Select}(C,T,\{\tilde{b}_i\}_i)]\!] & = & \{\texttt{Select}(C,T,b) \mid b \in [\![\tilde{b}_i]\!]\} \\
[\![\tilde{p}_1 \wedge \ldots \wedge \tilde{p}_n]\!] & = & \{p_1 \wedge \ldots \wedge p_n \mid p_j \in [\![\tilde{p}_j]\!]\} \\
[\![C = s]\!] & = & \{C = s\} \\
[\![C = \eta]\!] & = & \{C = e_t \mid e_t \in [\![\texttt{Progs}[\eta]]\!]\} \\
[\![C = \{s, \eta\}]\!] & = & [\![C = s]\!] \cup [\![C = \eta]\!]
\end{array}
$$
(c)

Figure 3: (a) The syntax of lookup transformation language $L_t$, (b) and (c) describe the syntax and semantics of data structure $D_t$ for succinctly representing a set of expressions from language $L_t$.

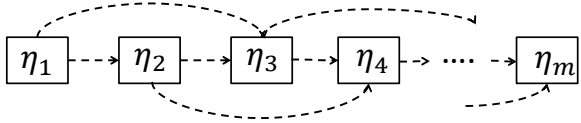

Figure 4: The reachability graph of nodes in Ex. 3.

makes use of the set of nodes $\{\eta_1, \ldots, \eta_m\}$ to succinctly represent $\Theta(2^m)$ transformations in $O(m)$ space.

**Exploiting CNF form of boolean conditions:** The second key aspect of our representation is exploiting the CNF form of boolean conditions to succinctly represent a huge set $\tilde{b}$ of conditionals. Consider a table $T$ with $n + 1$ columns $C_1, \ldots, C_{n+1}$, where the first $n$ columns constitute a primary key of the table and the table contains an entry $(s_1, s_2, \ldots, s_n, t)$. Consider the input-output example $(s_1, s_2, \ldots, s_m) \to t$ with $s_1 = s_2 = \cdots = s_{\max(m,n)}$. The number of transformations that are consistent with the given input-output example are $(m+1)^n$ because for indexing into each column $C_i$ of the table, we have $m + 1$ choices namely constant $s_1$ and the input string variables $v_1, \ldots, v_m$. This huge set of transformations can be represented succinctly in $O(n+m)$ space using our data structure as $(\{\eta_1, \eta_2\}, \eta_2, \texttt{Progs})$, where $\texttt{Progs}[\eta_1] = \{v_1, \ldots, v_m\}$, and $\texttt{Progs}[\eta_2] = \{\texttt{Select}(C_{n+1}, T, \tilde{b})\}$, $\tilde{b} = \bigwedge_{i=1}^{n}(C_i = \{s_1, \eta_1\})$.

THEOREM 1 (PROPERTIES OF DATA STRUCTURE $D_t$).
*(a) The number of transformations in $L_t$ that are consistent with a given example may be exponential in the number of reachable entries and number of columns in a candidate key. (b) However, the data structure $D_t$ can represent these potentially exponential number of transformations in polynomial size in number of reachable entries, number of candidate keys and number of columns in a candidate key.*

Proof of (a) follows from the two examples discussed above, while proof of (b) follows from Theorem 2(a).

### 4.3 Synthesis Algorithm for $L_t$

*Procedure* $\texttt{GenerateStr}_t$

The $\texttt{GenerateStr}_t$ procedure, shown in Figure 5(a), operates by iteratively computing a set of nodes $\tilde{\eta}$ and updating two maps $\texttt{Progs}$ and $\texttt{val}$ in the loop at Line 7. Each node $\eta \in \tilde{\eta}$ represents a string $\texttt{val}(\eta)$ that is present in some table entry. The inverse map $\texttt{val}^{-1}(a)$ returns the node corresponding to string $a$ or $\emptyset$ if no such node exist. The map $\texttt{Progs}$ associates every node $\eta$ to a set of expressions (of depth at most $\texttt{steps}$), each of which evaluates to string $\texttt{val}(\eta)$ on the input state $\sigma$. The key idea of the loop at Line 7 is to perform an iterative forward reachability analysis of the string values that can be generated in a single step (i.e., using a single $\texttt{Select}$ expression) from the string values computed in previous step, with the base case being the values of the input string variables.

Each iteration of the loop at Line 7 results in consideration of expressions whose depth is one larger than the set of expressions considered in the previous step. The depth of the expressions in language $L_t$ can be as much as the total number of entries in all of the relational tables combined. Since we have not observed any intended transformation that requires self-joins, we limit the depth consideration to a parameter $k$ whose value we set to be equal to the number of relational tables present in the spreadsheet. One might be tempted to use the predicate $(s \in \tilde{\eta} \vee \tilde{\eta}_{\texttt{old}} = \tilde{\eta})$ as a termination condition for the loop. However, this has two issues. The first issue is that it may happen co-incidentally that the output string $s$ is computable by a transformation of depth smaller than the depth of the intended transformation on a given example, and in that case we would fail to discover the correct transformation. The other major issue is that it might also happen that the intended transformation does not belong to the language $L_t$, in which case the search would fail, but only after consideration of all expressions whose depth is as large as the total number of entries in all relational tables combined together.

The generalized boolean condition $B$ is computed to be the set of all boolean conditions that uniquely identify row $r$ in table $T$ (Line 10). It considers the set of candidate keys of table $T$ and for each column $C'$ in a candidate key it learns the generalized predicate: $C' = \{T[C',r], \texttt{val}^{-1}[T[C',r]]\}$.

During the reachability computation, a node $\eta$ can be reached through multiple paths and therefore the set of expressions associated with the node ($\texttt{Progs}[\eta]$) needs to be updated accordingly. When a node is revisited, the algorithm computes the $\texttt{Select}$ expression with updated set of boolean conditions $B$ and adds it to the set (Line 15).

We now briefly describe how the $\texttt{GenerateStr}_t$ procedure computes the set of expressions for each node in Example 3. It first creates a node $\eta_1$ such that $\texttt{Progs}[\eta_1] = \{v_1\}$, $\texttt{val}(\eta_1) = s_1$ and the frontier of reachable nodes is set as $\tilde{\eta}_{\texttt{diff}} = \{\eta_1\}$. We use node $\eta_i$ to denote the node corresponding to string $s_i$ such that $\texttt{val}(\eta_i) = s_i$. The algorithm then finds that the table entry $T_1[C_1, 1]$ is reachable from node $\eta_1$ with the generalized boolean condition $B = \{C_1 = \{s_1, v_1\}\}$. The algorithm then makes the other column entries in the row, namely $T_1[C_2, 1]$ and $T_1[C_3, 1]$

```
GenerateStr_t(σ: Input state, s: Output string)
1  η̃ := ∅; η̃_old := ∅; steps := 0;
2  foreach input variable v_i:
3    if ((η := val^{-1}(σ(v_i))) = ⊥)
4    then { η := NewNode(); η̃ := η̃ ∪ {η};
5           val(η) := σ(v_i); Progs[η] := ∅; }
6    Progs[η] := Progs[η] ∪ {v_i};
7  while (steps++ ≤ k ∧ η̃_old ≠ η̃)
8    η̃_diff := η̃ − η̃_old; η̃_old := η̃;
9    foreach table T, col C, row r s.t.
          T[C,r] = val(η) for some η ∈ η̃_diff
10     B := { ⋀_{C'∈cKey} (C' = {T[C',r], val^{-1}(T[C',r])}) |
              cKey ∈ CandidateKeys(T)};
11     foreach column C' of table T s.t. C' ≠ C:
12       if ((η := val^{-1}(T[C',r])) = ⊥)
13       then { η := NewNode(); η̃ := η̃ ∪ {η};
14              val(η) := T[C',r]; Progs[η] := ∅; }
15       Progs[η] := Progs[η] ∪ {Select(C',T,B)};
16 return (η̃, val^{-1}(s), Progs);
              (a)
```

$$\texttt{Intersect}_t((\tilde{\eta}_1, \eta_1^t, \texttt{Progs}_1), = (\tilde{\eta}_1 \times \tilde{\eta}_2, (\eta_1^t, \eta_2^t), \texttt{Progs}_{12})$$
$$(\tilde{\eta}_2, \eta_2^t, \texttt{Progs}_2))$$
$$\text{where } \texttt{Progs}_{12}[(\eta_1, \eta_2)] = \texttt{Intersect}_t(\texttt{Progs}_1[\eta_1], \texttt{Progs}_2[\eta_2])$$
$$\texttt{Intersect}_t(v_i, v_i) = v_i$$
$$\texttt{Intersect}_t(\texttt{Select}(C, T, B), = \texttt{Select}(C, T, B'')$$
$$\texttt{Select}(C, T, B'))$$
$$\text{where } B'' = \texttt{Intersect}_t(B, B')$$
$$\texttt{Intersect}_t(\{\tilde{b}_i\}_i, \{\tilde{b}'_i\}_i) = \{\texttt{Intersect}_t(\tilde{b}_i, \tilde{b}'_i)\}_i$$
$$\texttt{Intersect}_t(\{\tilde{p}_i\}_i, \{\tilde{p}'_i\}_i) = \{\texttt{Intersect}_t(\tilde{p}_i, \tilde{p}'_i)\}_i$$
$$\texttt{Intersect}_t(C = s, C = s) = C = s$$
$$\texttt{Intersect}_t(C = \eta_1, C = \eta_2) = C = (\eta_1, \eta_2)$$
$$\texttt{Intersect}_t(C = \{s, \eta_1\}, = C = \{s, (\eta_1, \eta_2)\}$$
$$C = \{s, \eta_2\})$$
$$\texttt{Intersect}_t(C = \{s_1, \eta_1\}, = C = \{(\eta_1, \eta_2)\}, \text{ if } s_1 \neq s_2$$
$$C = \{s_2, \eta_2\})$$
(b)

Figure 5: (a) The GenerateStr$_t$ procedure for generating the set of all expressions (of depth at most $k$) in language $L_t$ that are consistent with a given input-output example. (b) The Intersect$_t$ procedure for intersecting sets of expressions from language $L_t$. The Intersect$_t$ procedure returns $\emptyset$ in all other cases.

also reachable and creates nodes $\eta_2$ and $\eta_3$ corresponding to them, and then sets $\texttt{Progs}[\eta_2] = \{\texttt{Select}(C_2, T_1, B)\}$ and $\texttt{Progs}[\eta_3] = \{\texttt{Select}(C_3, T_1, B)\}$. In the next iteration of loop, the frontier of reachable set is updated to $\tilde{\eta}_{\text{diff}} = \{\eta_2, \eta_3\}$ and the nodes that are reachable from this set are next computed. The algorithm finds that table entry $T_2[C_1, 1]$ is reachable from node $\eta_2$ and thereby makes nodes $\eta_3$ and $\eta_4$ reachable as well with corresponding Select expressions. Similarly, the nodes $\eta_4$ and $\eta_5$ become reachable from $\eta_3$. In this manner, the algorithm keeps computing the set of reachable table entries iteratively until $k$ iterations, where $k$ is set to the number of relational tables $n$.

*Procedure* Intersect$_t$

The Intersect$_t$ procedure takes two sets of expressions in $L_t$ as input and computes the set of expressions that are common to both the sets. (Both input and output sets are represented using the data structure $D_t$.) Figure 5(b) describes the Intersect$_t$ procedure for intersecting the sets of $L_t$ expressions using a set of rules that are pattern matched for execution. For intersecting two expressions $(\tilde{\eta}_1, \eta_1^t, \texttt{Progs}_1)$ and $(\tilde{\eta}_2, \eta_2^t, \texttt{Progs}_2)$, we take the cross product of the set of nodes to get the new set of nodes $\tilde{\eta}_{12} = (\tilde{\eta}_1 \times \tilde{\eta}_2)$ with the target node $(\eta_1^t, \eta_2^t)$, and compute the new $\texttt{Progs}_{12}$ map for each node $(\eta_1, \eta_2) \in \tilde{\eta}_{12}$ by intersecting their corresponding maps $\texttt{Progs}_1[\eta_1]$ and $\texttt{Progs}_2[\eta_2]$ respectively. The intersect rule for two select expressions requires the column name and table id to be the same and intersects the conditionals recursively. The candidate keys $\tilde{b}_i$ as well as each column conditional $\tilde{p}$ in a candidate key are intersected individually maintaining their corresponding orderings.

THEOREM 2 (SYNTHESIS ALGORITHM PROPERTIES). *(a) The procedure* GenerateStr$_t$ *is sound and k-complete. The computational complexity of* GenerateStr$_t$ *procedure (and hence the size of the data structure constructed by it) is $O(t^2 p m)$ where $t$ is the number of reachable strings in $k$ iterations, $p$ is the maximum number of candidate keys in any table and $m$ is the maximum number of columns in any candidate key. (b) The procedure* Intersect$_t$ *is sound and complete. The computational complexity of* Intersect$_t$ *procedure (and hence the size of the data structure returned by it) is $O(d^2)$, where $d$ is the size of the input data structures.*

The proof of Theorem 2 is given in [16].

## 4.4 Ranking

In this section, we define a partial order between expressions in $L_t$ that we use for ranking of these expressions. We prefer expressions of smaller depth (fewer nested chains of Select expressions). We prefer lookup expressions that use distinct tables for join queries (the most common scenario for end-users) as opposed to expressions involving self-joins. We prefer conditionals that consist of fewer predicates and prefer predicates that involve comparing columns with other table entries or input variables (as opposed to comparing columns with constant strings).

## 5. SEMANTIC TRANSFORMATIONS

We now present an extension of the lookup transformation language $L_t$ (described in §4) with a syntactic transformation language $L_s$ (from [8]). This extended language $L_u$, also referred to as semantic transformation language, adds two key capabilities to $L_t$: (i) It allows for lookup transformations that involve performing syntactic manipulations (such as substring, concatenation, etc.) on strings before using them to perform lookups, and (ii) It allows for performing syntactic manipulations on lookup outputs (which can then be used for performing further lookups or for generating the output string). This extension, as we show in §6, also enables us to model transformations on strings representing standard data types such as date, time, etc. We first describe a syntactic transformation language.

*Syntactic Transformation Language $L_s$ (Background).*
Gulwani [8] introduced an expression language for performing syntactic string transformations. We reproduce here a small subset of (the rules of) that language and call it $L_s$ (with $e_s$ being the top-level symbol).

$$e_s := \texttt{Concatenate}(f_1, \ldots, f_n) \mid f$$
$$\text{Atomic expr f} := \texttt{ConstStr}(s) \mid v_i \mid \texttt{SubStr}(v_i, p_1, p_2)$$
$$\text{Position p} := k \mid \texttt{pos}(r_1, r_2, c)$$
$$\text{Integer expr c} := k \mid k_1 w + k_2$$
$$\text{Regular expr r} := \epsilon \mid \tau \mid \texttt{TokenSeq}(\tau_1, \ldots, \tau_n)$$

The formal semantics of $L_s$ can be found in [8]. For completeness, we briefly describe some key aspects of this language. The top-level expression $e_s$ is either an atomic expression f or is obtained by concatenating atomic expressions $f_1, \ldots, f_n$ using the Concatenate constructor. Each atomic expression f can either be a constant string $\texttt{ConstStr}(s)$, an input string variable $v_i$, or a substring of some input string $v_i$. The substring expression $\texttt{SubStr}(v_i, p_1, p_2)$ is defined partly by two *position expressions* $p_1$ and $p_2$, each of which implicitly refers to the (subject) string $v_i$ and must evaluate to a position within the string $v_i$. (A string with $\ell$ characters has $\ell + 1$ positions, numbered from 0 to $\ell$ starting from left.) $\texttt{SubStr}(v_i, p_1, p_2)$ is the substring of string $v_i$ in between positions $p_1$ and $p_2$. A position expression represented by a non-negative constant $k$ denotes the $k^{th}$ position in the string. For a negative constant $k$, it denotes the $(\ell + 1 + k)^{th}$ position in the string, where $\ell = \texttt{Length}(s)$. A position expression $\texttt{pos}(r_1, r_2, c)$ evaluates to a position $t$ in the subject string $s$ such that regular expression $r_1$ matches some suffix of $s[0:t]$, and $r_2$ matches some prefix of $s[t:\ell]$, where $\ell = \texttt{Length}(s)$. Furthermore, if c is positive (negative), then $t$ is the $|c|^{th}$ such match starting from the left side (right side). We use the expression $s[t_1 : t_2]$ to denote the substring of $s$ between positions $t_1$ and $t_2$. We use notation $\texttt{SubStr2}(v_i, r, c)$ as an abbreviation to denote the $c^{th}$ occurrence of r in $v_i$, i.e., $\texttt{SubStr}(v_i, \texttt{pos}(\epsilon, r, c), \texttt{pos}(r, \epsilon, c))$.

A regular expression r is either a token $\tau$, a token sequence $\texttt{TokenSeq}(\tau_1, \ldots, \tau_n)$ or $\epsilon$ (which matches the empty string). The tokens $\tau$ range over some finite (but easily extensible) set and typically correspond to character classes and special characters. For example, tokens UpperTok, NumTok, and AlphTok match a nonempty sequence of uppercase alphabetic characters, numeric digits, and alphanumeric characters respectively. DecNumTok matches a nonempty sequence of numeric digits and/or decimal point. SlashTok matches the slash character. Special tokens StartTok and EndTok match the beginning and end of a string respectively.

EXAMPLE 4. *An Excel user wanted to transform names to a format where the last name is followed by the initial letter of the first name, e.g., "Alan Turing" → "Turing A". An expression in $L_s$ that can perform this transformation is: $\texttt{Concatenate}(f_1, \texttt{ConstStr}(\text{" "}), f_2)$ where $f_1 \equiv \texttt{SubStr2}(v_1, \texttt{AlphTok}, 2)$ and $f_2 \equiv \texttt{SubStr2}(v_1, \texttt{UpperTok}, 1)$. This expression constructs the output sting by concatenating the $2^{nd}$ word of input string, the constant string whitespace, and the $1^{st}$ capital letter in input string.*

For more details on the syntactic transformation language $L_s$, see [8]. Now we present the extended language $L_u$.

## 5.1 Semantic Transformation Language $L_u$

Let $R_t$ and $R_s$ denote the set of grammar rules of languages $L_t$ and $L_s$ respectively. We subscript each non-terminal in the two languages with $t$ and $s$ for disambiguating the names of non-terminals in the extended language.

For example, $f_s$ denotes the atomic expression f of the syntactic transformation language $L_s$. The expression grammar of the extended language $L_u$ consists of rules $R_t \cup R_s$ in which the following rules are modified (with modifications shown in bold), and with $e_s$ as the top-level symbol.

$$\text{Atomic expr } f_s := \texttt{ConstStr}(s) \mid \mathbf{e_t} \mid \texttt{SubStr}(\mathbf{e_t}, p_{s_1}, p_{s_2})$$
$$\text{Predicate } p_t := C = s \mid C = \mathbf{e_s}$$
$$e_s := \texttt{Concatenate}(f_{s_1}, \ldots, f_{s_n}) \mid f_s$$

The top-level expression $e_s$ of the extended language is either an atomic expression $f_s$ or a Concatenate operation on a sequence of atomic expressions $f_{s_i}$, as before. However, the atomic expression $f_s$ is updated to consist of a lookup expression $e_t$ or its substring $\texttt{SubStr}(e_t, p_{s_1}, p_{s_2})$ (as opposed to only an input variable $v_i$ or its substring). This lets the language model transformations that perform syntactic manipulations over table lookup outputs. The other modification is in the predicate expression $p_t$ of language $L_t$, where we modify the conditional expression $C = e_t$ to $C = e_s$. This enables the language to model lookup transformations that perform syntactic manipulations on strings before performing the lookup. The updated rules have expected semantics and can be defined in a similar fashion as the semantics of rules in $L_t$ and $L_s$. We now illustrate the expressiveness of the extended language using few examples.

The transformation in Example 1 can be represented in $L_u$ as: $\texttt{Concatenate}(f_1, \texttt{ConstStr}(\text{``+0.''}), f_2, \texttt{ConstStr}(\text{``*''}), f_3)$,
$f_1 \equiv \texttt{Select}(\texttt{Price}, \texttt{CostRec}, \texttt{Id} = f_4 \wedge \texttt{Date} = f_5)$,
$f_4 \equiv \texttt{Select}(\texttt{Id}, \texttt{MarkupRec}, \texttt{Name} = v_1)$,
$f_5 \equiv \texttt{SubStr}(v_2, \texttt{pos}(\texttt{SlashTok}, \epsilon, 1), \texttt{pos}(\texttt{EndTok}, \epsilon, 1))$,
$f_2 \equiv \texttt{SubStr2}(f_6, \texttt{NumTok}, 1)$, $f_3 \equiv \texttt{SubStr2}(f_1, \texttt{DecNumTok}, 1)$,
$f_6 \equiv \texttt{Select}(\texttt{Markup}, \texttt{MarkupRec}, \texttt{Name} = v_1)$.

The expression $f_4$ looks up the Id of the item in $v_1$ from the MarkupRec table and expression $f_5$ generates a substring of the date in $v_2$, which are then used to look up the Price of the item from the CostRec table ($f_1$). The expression $f_6$ looks up the Markup percentage of the item from the MarkupRec table and $f_2$ generates a substring of this lookup value by extracting the first numeric token (thus removing the % sign). Similarly, the expression $f_3$ generates a substring of $f_1$, removing the $ sign. Finally, the top-level expression concatenates $f_1$, $f_2$, and $f_3$ with constant strings "+0." and "*".

EXAMPLE 5. **Indexing with concatenated strings:** *A bike merchant maintained an inventory of BikePrices table, and wanted to compute the price quote table by performing lookup of bike Price after concatenating the bike name ($v_1$) and the engine cc value ($v_2$) as shown in Figure 6.*

| Input $v_1$ | Input $v_2$ | Output |
|---|---|---|
| Honda | 125 | 11,500 |
| Ducati | 100 | **10,000** |
| Honda | 250 | **19,000** |
| Ducati | 250 | **18,000** |

| BikePrices ||
|---|---|
| Bike | Price |
| Ducati100 | 10,000 |
| Ducati125 | 12,500 |
| Ducati250 | 18,000 |
| Honda125 | 11,500 |
| Honda250 | 19,000 |
| … | … |

**Figure 6: A lookup transformation that requires concatenating input strings before performing selection from a table.**

The desired transformation can be expressed in the extended language as: $\texttt{Select}(\texttt{Price}, \texttt{BikePrices}, \texttt{Bike} = e_s)$ where $e_s = \texttt{Concatenate}(v_1, v_2)$. The expression $e_s$ concatenates the two input string variables $v_1$ and $v_2$, which is then used to perform the lookup in the BikePrices table.

EXAMPLE 6. **Concatenating table outputs:** *A user had a series of three company codes in a column and wanted to expand them into the corresponding series of company names using a table* **Comp** *that mapped company codes to the company names as shown in Figure 7.*

| Input $v_1$ | Output |
|---|---|
| c4 c3 c1 | Facebook Apple Microsoft |
| c2 c5 c6 | **Google IBM Xerox** |
| c1 c5 c4 | **Microsoft IBM Facebook** |
| c2 c3 c4 | **Google Apple Facebook** |

| Comp | |
|---|---|
| Id | Name |
| c1 | Microsoft |
| c2 | Google |
| c3 | Apple |
| c4 | Facebook |
| c5 | IBM |
| c6 | Xerox |
| ... | ... |

**Figure 7: A nested syntactic and lookup transformation. It requires concatenating results of multiple lookup transformations, each of which involves a selection operation that indexes on some substring of the input.**

This transformation is expressed in $L_u$ as:
$\texttt{Concatenate}(\texttt{f}_1, \texttt{ConstStr}(\text{" "}), \texttt{f}_2, \texttt{ConstStr}(\text{" "}), \texttt{f}_3)$, where $\texttt{f}_1 \equiv \texttt{Select}(\texttt{Name}, \texttt{Comp}, \texttt{Id} = \texttt{SubStr2}(v_1, \texttt{AlphTok}, 1))$, $\texttt{f}_2 \equiv \texttt{Select}(\texttt{Name}, \texttt{Comp}, \texttt{Id} = \texttt{SubStr2}(v_1, \texttt{AlphTok}, 2))$ and $\texttt{f}_3 \equiv \texttt{Select}(\texttt{Name}, \texttt{Comp}, \texttt{Id} = \texttt{SubStr2}(v_1, \texttt{AlphTok}, 3))$. The expressions $\texttt{f}_1$, $\texttt{f}_2$, and $\texttt{f}_3$ extract the first, second, and third words from the input string respectively, which are then used for performing the table lookups and the results are concatenated with whitespaces to obtain the output string.

## 5.2 Data Structure for Set of Expressions in $L_u$

Let $\tilde{R}_t$ and $\tilde{R}_s$ denote the set of grammar rules for the data structures that represent set of expressions in languages $L_t$ and $L_s$ respectively (See [8] for description of $\tilde{R}_s$). We construct the grammar rules for the data structure that represents set of expressions in the extended language $L_u$ by taking the union of the two rules $\tilde{R}_t \cup \tilde{R}_s$ and modifying some rules as follows:

$$\tilde{f}_s ::= \texttt{ConstStr}(s) \mid \tilde{\mathbf{e}}_\mathbf{t} \mid \texttt{SubStr}(\tilde{\mathbf{e}}_\mathbf{t}, \tilde{p}_{\mathbf{s}_1}, \tilde{p}_{\mathbf{s}_2})$$
$$\tilde{p}_t ::= C = s \mid \mathbf{C} = \tilde{\mathbf{e}}_\mathbf{s}$$
$$\tilde{e}_s ::= \texttt{Dag}(\tilde{\alpha}, \alpha^\mathbf{s}, \alpha^\mathbf{t}, \tilde{\xi}, \mathbf{W}) \mid \tilde{f}_s, \text{ where } W : \tilde{\xi} \to 2^{\mathbf{f}_s}$$

The most interesting aspect of this data structure is the $\texttt{Dag}(\tilde{\alpha}, \alpha^s, \alpha^t, \tilde{\xi}, W)$ construct, which succinctly represents a set of Concatenate expressions in $L_u$ using a dag (directed acyclic graph), where $\tilde{\alpha}$ is a set of nodes containing two distinct source and target nodes $\alpha^s$ and $\alpha^t$, $\tilde{\xi}$ is a set of edges over nodes in $\tilde{\alpha}$ that induces a dag, and $W$ maps each edge in $\tilde{\xi}$ to a set of atomic expressions. The semantics $[\![.]\!]$ of the Dag constructor is:

$[\![\texttt{Dag}(\tilde{\alpha}, \alpha^s, \alpha^t, \tilde{\xi}, W)]\!] = \{\texttt{Concatenate}(\texttt{f}_{s_1}, \ldots, \texttt{f}_{s_n}) \mid$
$\texttt{f}_{s_i} \in [\![W(\tilde{\xi}_i)]\!], \xi_1, .., \xi_n \in \tilde{\xi} \text{ form a path between } \alpha^s \text{ and } \alpha^t\}$

The set of all Concatenate expressions represented by the Dag constructor includes exactly those whose ordered arguments belong to the corresponding edges on any path from $\alpha^s$ to $\alpha^t$. This dag representation is similar to the representation of string expressions in [8]. However, in our case, the edges of the dag consist of more sophisticated (substrings of) lookup expressions, whose predicates can in-turn be represented using nested-dags.

Consider Example 6, where the input string is "c4 c3 c1" and the output string is "Facebook Apple Microsoft" (of length 24). The dag $G$ that represents the set of all transformations consistent with this input-output pair is shown in Figure 8. For better readability, we only show some of the relevant nodes and edges of the dag $G$. The edge from node 0 to node 8 corresponds to all expressions $\tilde{e}_1$ that generate the string Facebook. One of the lookup transformations, $\texttt{Select}(\texttt{Name}, \texttt{Comp}, \texttt{Id} = \tilde{f}_1)$, in $\tilde{e}_1$ requires syntactic transformations $\tilde{f}_1$ to extract substring $c1$ from the input string, where $\tilde{f}_1$ is itself represented as a nested-dag as shown in the figure. The edges for expressions $\tilde{e}_3$ and $\tilde{e}_5$ also consist of similar nested-dags.

THEOREM 3 (PROPERTIES OF DATA STRUCTURE $D_u$).
*(a) The number of transformations in $L_u$ that are consistent with a given input-output example may be exponential in the number of reachable entries, the number of columns in a primary key, and the length of the largest reachable string.*
*(b) However, the data structure $D_u$ can represent these potentially exponential number of transformations in size polynomial in the number of reachable entries, the number of primary keys, the number of columns in a primary key, and the length of the largest reachable string.*

(a) Proof of (a) follows from Theorem 1(a). The number of transformations can also be exponential in the length of the largest reachable string as we are using $\texttt{GenerateStr}_s$ procedure for checking reachability which has this worst case complexity. (b) We show that the size of the data structure generated is $O(t^2 \, p \, m \, \ell^2)$ in Theorem 4(a).

## 5.3 Synthesis Algorithm for $L_u$

*Procedure* $\texttt{GenerateStr}_u$

Recall that the $\texttt{GenerateStr}_t$ procedure for language $L_t$ (§4.3) performs reachability on table entries based on exact string matches $(T[C, r] = \eta)$. The key idea in case of the language $L_u$ is to perform a more relaxed reachability on table entries taking into account the possibility of performing syntactic manipulations on previously reachable strings.

We first define a $\texttt{GenerateStr}'_t$ procedure by making two modifications to the $\texttt{GenerateStr}_t$ procedure. First, we replace the condition "$T[C, r] = \texttt{val}(\eta)$" in $\texttt{GenerateStr}_t$ (Line 9 in Figure 5(a)) by the condition "$\texttt{GenerateStr}_s(\sigma \cup \tilde{\eta}, T[C, r])$ contains any expression that uses a variable from $\sigma \cup \tilde{\eta}$". We use the notation $\sigma \cup \tilde{\eta}$ to denote a state that extends $\sigma$ and maps $\eta$ to $\texttt{val}(\eta)$ for all $\eta \in \tilde{\eta}$. The $\texttt{GenerateStr}'_t$ procedure marks a table entry as reachable if it can be computed using the $\texttt{GenerateStr}_s$ procedure (from [8]) on previously reachable strings. The $\texttt{GenerateStr}_s$ procedure can perform concatenation of constant strings and substrings of previously reachable strings ($\tilde{\eta}$). We add an additional check that the expressions synthesized by $\texttt{GenerateStr}_s$ contains a variable from $\sigma \cup \tilde{\eta}$ to avoid expressions containing only constant string expressions. For our experiments, we enforce an even stronger restriction that there exists a string $\eta \in (\sigma \cup \tilde{\eta})$ such that either $T[C, r]$ is substring of $\eta$ or $\eta$ is a substring of $T[C, r]$. This provides efficiency without any practical loss of precision. The second modification is in Line 10 in Figure 5(a), where we replace the generalized predicate with $C' = \{\texttt{GenerateStr}_s(\sigma \cup \tilde{\eta}, T[C', r])\}$.

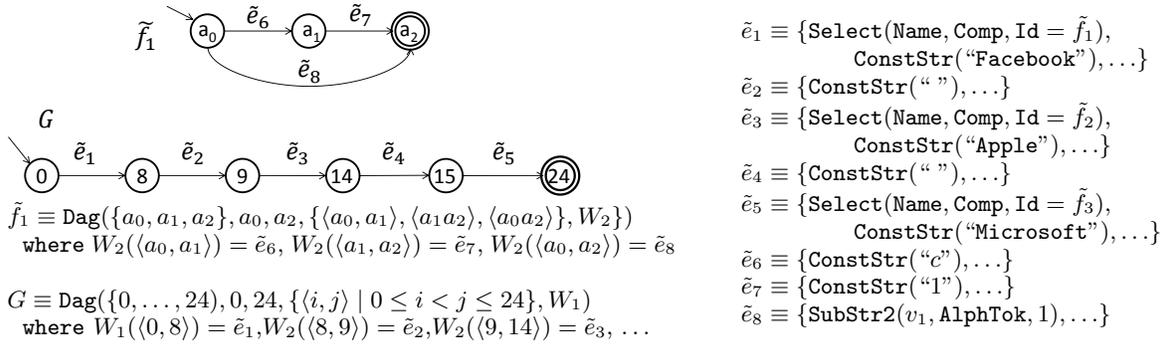

**Figure 8: A partial Dag representation of the set of expressions in Example 6.**

The $\text{GenerateStr}_u$ procedure for the extended language $L_u$ can now be defined as:

$\underline{\text{GenerateStr}_u}(\sigma: \text{ Input state}, s: \text{ Output string})$
1 $(\tilde{\eta}, \eta^t, \text{Progs}) = \text{GenerateStr}'_t(\sigma, s);$
2 return $\text{GenerateStr}_s(\sigma \cup \tilde{\eta}, s);$

The $\text{GenerateStr}_u$ procedure first constructs the set of all reachable table entries ($\tilde{\eta}$) from the set of input strings in $\sigma$ using the $\text{GenerateStr}'_t$ procedure. It then uses the $\text{GenerateStr}_s$ procedure to construct the Dag for generating the output string $s$ from the set of strings that includes values of input variables (in state $\sigma$) as well as the reachable table entries (represented by $\tilde{\eta}$).

Consider the first input-output pair ("c4 c3 c1", "Facebook Apple Microsoft") in Example 6. The $\text{GenerateStr}_u$ algorithm first uses the $\text{GenerateStr}'_t$ procedure to compute the set of reachable table entries. The $\text{GenerateStr}'_t$ procedure finds that the table entry $\text{T[Id,4]} = \text{c4}$ is reachable from the input string "c4 c3 c1" as there exists an expression $\text{SubStr2}(v_1, \text{AlphTok}, 1)$ that can generate the string c4. It then adds the string Facebook from row 4 to the reachable set $\tilde{\eta}$. Similarly, table entries c3, Apple, c1, and Microsoft are also added to $\tilde{\eta}$. It then uses the $\text{GenerateStr}_s$ procedure to construct a dag for generating the output string "Facebook Apple Microsoft" from the set {"c4 c3 c1", c1, c2, c3, Facebook, Apple, Microsoft} ($\sigma \cup \tilde{\eta}$). It first creates a Dag of 25 nodes ($\tilde{\alpha} = \{0, \ldots, 24\}$) with 0 as the source node and 24 as the target node. The algorithm then assigns a set of expressions to each edge $\langle i, j \rangle$ that can generate the substring $s[i,j]$. For example, the algorithm adds the expression that selects the string Facebook from $\sigma \cup \tilde{\eta}$ to the edge $\langle 0, 8 \rangle$; the expression in turn corresponds to a lookup transformation with nested sub-dags as shown in Figure 8. Similarly, it adds the expression $\tilde{e}_2$ that generates a whitespace to the edge $\langle 8, 9 \rangle$, and so on.

*Procedure* $\text{Intersect}_u$

The $\text{Intersect}_u$ procedure for the extended language $L_u$ consists of the union of rules of $\text{Intersect}_t$ and $\text{Intersect}_s$ procedures along with the following four additional rules.

$\text{Intersect}_u(\tilde{e}_t, \tilde{e}'_t) = \text{Intersect}_t(\tilde{e}_t, \tilde{e}'_t)$
$\text{Intersect}_u(C = \tilde{e}_s, C = \tilde{e}'_s) = (C = \text{Intersect}_s(\tilde{e}_s, \tilde{e}'_s))$
$\text{Intersect}_u(\text{SubStr}(\tilde{e}_t, \tilde{p}_{s_1}, \tilde{p}_{s_2}), \text{SubStr}(\tilde{e}'_t, \tilde{p}'_{s_1}, \tilde{p}'_{s_2})) =$
$\quad \text{SubStr}(\text{Intersect}_t(\tilde{e}_t, \tilde{e}'_t), \text{IntersectPos}(\tilde{p}_{s_1}, \tilde{p}'_{s_1}),$
$\quad\quad \text{IntersectPos}(\tilde{p}_{s_2}, \tilde{p}'_{s_2}))$

$\text{Intersect}_u(\text{Dag}(\tilde{\alpha}_1, \alpha^s_1, \alpha^t_1, \tilde{\xi}_1, W_1), \text{Dag}(\tilde{\alpha}_2, \alpha^s_2, \alpha^t_2, \tilde{\xi}_2, W_2))$
$= \text{Dag}(\tilde{\alpha}_1 \times \tilde{\alpha}_2, (\alpha^s_1, \alpha^s_2), (\alpha^t_1, \alpha^t_2), \tilde{\xi}_{12}, W_{12})$, where
$\tilde{\xi}_{12} = \{\langle (\alpha_1, \alpha_2), (\alpha'_1, \alpha'_2) \rangle \mid \langle \alpha_1, \alpha'_1 \rangle \in \tilde{\xi}_1, \langle \alpha_2, \alpha'_2 \rangle \in \tilde{\xi}_2\}$
and $W_{12}(\langle (\alpha_1, \alpha_2), (\alpha'_1, \alpha'_2) \rangle) = \{\text{Intersect}_s(\text{f}, \text{f}') \mid$
$\quad \text{f} \in W_1(\langle \alpha_1, \alpha'_1 \rangle), \text{f}' \in W_2(\langle \alpha_2, \alpha'_2 \rangle)\}$

The intersect rule for Dag intersects the two dags in a manner similar to the intersection of two finite state automatons. The new mapping $W_{12}$ is computed by performing intersection of the expressions on the two corresponding edges of the dags. (Rules for $\text{Intersect}_s$ are defined in [8].)

THEOREM 4 (SYNTHESIS ALGORITHM PROPERTIES). *(a) The $\text{GenerateStr}_u$ procedure is sound and k-complete. The computational complexity of $\text{GenerateStr}_u$ procedure is $O(t^2\, p\, m\, \ell^4)$ (assuming $O(l^2)$ complexity for the new check on Line 9), and the size of the data structure constructed by it is $O(t^2\, p\, m\, \ell^2)$, where $t$ is the number of reachable strings in k iterations, $p$ is the maximum number of candidate keys of any table, $m$ is the maximum number of columns in any candidate key, and $\ell$ is the length of the longest reachable string. (b) The $\text{Intersect}_u$ procedure is sound and complete. The computational complexity of $\text{Intersect}_u$ (and hence the size of the data structure returned by it) is $O(d^2)$, where $d$ is the size of the input data structures.*

The proof of Th. 4 and more details about $O(l^2)$ assumption for check on Line 9 are given in [16]. The worst-case quadratic blowup in the size of the output returned by the Intersect procedure doesn't happen in practice (as we report in §7) making the synthesis algorithm very efficient.

### 5.4 Ranking

The partial orders of ranking schemes of $L_t$ and $L_s$ are also used to rank expressions in $L_u$. In addition, we define some additional partial orders for expressions in $L_u$. We prefer lookup expressions that match longer strings in table entries for indexing than the ones that match shorter strings. We prefer lookup expressions with fewer constant string expressions and ones that generate longer output strings.

## 6. STANDARD DATA TYPES

The language $L_u$ can also model a rich class of transformations on strings that represent standard data types such as date, time, phone numbers, currency, or addresses. Manipulation of these data types typically requires some background knowledge associated with these data types. For example, for dates we have the knowledge that month 2 corresponds to the string February, or for phone numbers we have the knowledge that 90 is the ISD code for Turkey. This background knowledge can be encoded as a set of relational tables in our framework (once and for all). This allows the synthesis algorithm for $L_u$ to learn transformations over strings representing these data types. We now present some examples from Excel help-forums where users were struggling with performing manipulations over such strings.

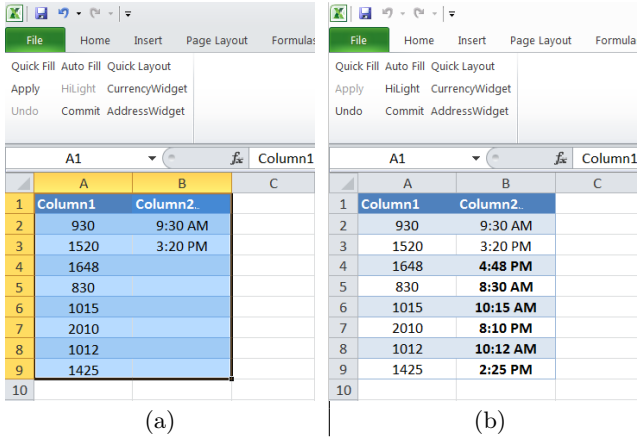

Figure 9: User interface of our programming-by-example Excel add-in. (a) and (b) are the screenshots before and after clicking the Apply button.

| Input $v_1$ | Output |
|---|---|
| 6-3-2008 | Jun 3rd, 2008 |
| 3-26-2010 | **Mar 26th, 2010** |
| 8-1-2009 | **Aug 1st, 2009** |
| 9-24-2007 | **Sep 24th, 2007** |

Figure 10: Formatting dates using examples.

EXAMPLE 7 (TIME MANIPULATION). *An Excel user needed to have* spot times *(shown in column 1 in Figure 9(a)) converted to the hh:mm AM/PM format (as shown in column 2 in Figure 9(a)). The Excel user posted this problem on a help-forum to which an expert responded by providing the following macro: TEXT(C1,"00 00")+0 When we showed this example to a team of Excel experts in a live presentation, it drew a response: "There are 40 different ways of doing this!". When we asked them to describe any one of those ways, we got the response "I don't exactly remember how to do it. I will have to investigate".*

We encode the background knowledge concerning time in a table Time with three columns 24Hour, 12Hour and AMPM, where the first column constitutes a primary key, and so does the combination of the second and third columns. The table is populated with 24 entries: $(0, 0, AM)$, ..., $(11, 11, AM)$, $(12, 12, PM)$, $(13, 1, PM)$, ..., $(23, 11, PM)$. The desired transformation can be represented in our language as:
Concatenate(Select(12Hour, Time, $b_1$), ConstStr(":"),
  SubStr($v_1$, −3, −1), ConstStr(" "), Select(AMPM, Time, $b_1$))
where $b_1 \equiv (\text{24Hour} = \text{SubStr}(v_1, \text{pos}(\text{StartTok}, \epsilon, 1), -3))$.

The SubStr expression in $b_1$ computes the substring of the input between the start and $3^{rd}$ character from end, to compute the hour part of the time in column $v_1$. This hour string is then used to perform lookup in table Time to compute its corresponding 12Hour format and AMPM value. These lookup strings are then concatenated with the minute part of the input string and constant strings : and whitespace.

EXAMPLE 8 (DATE MANIPULATION). *An Excel user wanted to convert dates from one format to another as shown in Figure 10, and the fixed set of hard-coded date formats supported by Excel 2010 do not match the input and output formats. Thus, the user solicited help on a forum.*

We encode the background knowledge concerning months in a table Month with two columns MN and MW, where each of the columns constitutes a candidate key by itself. The table is populated with 12 entries: $(1, \text{January})$, ..., $(12, \text{December})$. We also maintain a table DateOrd with two columns Num and Ord, where the first column constitutes a primary key. The table contains 31 entries $(1, \text{st}), (2, \text{nd}), (3, \text{rd}), \ldots, (31, \text{st})$. The desired transformation is represented in $L_u$ as:
Concatenate(SubStr(Select(MW, Month, MN = $e_1$),
   pos(StartTok, $\epsilon$, 1), 3), ConstStr(" "), $e_2$,
   Select(Ord, DateOrd, Num = $e_2$), ConstStr(", "), $e_3$)
where $e_1 = \text{SubStr2}(v_1, \text{NumTok}, 1)$, $e_2 = \text{SubStr2}(v_1, \text{NumTok}, 2)$, and $e_3 = \text{SubStr2}(v_1, \text{NumTok}, 3)$.

The expression concatenates the following strings: the string obtained by lookup of first number token of $v_1$ in table Month, the constant string whitespace, the second number token of $v_1$, the string obtained by lookup of second number token of $v_1$ in table DateOrd, the constant string ", ", and the string corresponding to third number token of $v_1$.

Unfortunately, it is not possible to encode semantics of data-types with infinite domains using relational tables. One such data-type is numbers, which often entail rounding and formatting transformations [17].

## 7. EXPERIMENTS

We have implemented the inductive synthesis algorithm for the transformation language $L_u$ in C# as an add-in for Microsoft Excel Spreadsheet system as shown in Figure 9. We hard-code a few useful relational tables of our own (such as the one that maps month numbers to month names), while also allowing the user to point to existing Excel tables to be used for performing the transformation.

**Benchmarks:** We report experimental results on a set of 50 problems collected from several Excel help-forums and the Excel product team (including all problems described in this paper). Out of these 50 problems, 12 problems can be modeled in the lookup language $L_t$ whereas the remaining 38 of them require the extended language $L_u$. The detailed description of these 50 problems can be found in [16].

**Effectiveness of data structure $D_u$:** We first present the statistics about the number of expressions in $L_u$ that are consistent with the user-provided set of input-output examples for each benchmark problem in Figure 11(a). The figure shows that the number of such consistent expressions are very large and are typically in the range from $10^{10}$ to $10^{30}$. Figure 11(b) shows that the size of our data structure $D_u$ to represent this large number of expressions typically varies from 100 to 2000, where each terminal symbol in the grammar rules of the data structure contributes a unit size to the size of the data structure.

**Effectiveness of ranking:** Use of a ranking scheme enables users to provide fewer input-output examples to automate their repetitive task. Hence, the effectiveness of our ranking scheme can be measured by the number of examples required for the intended program to be ranked as the topmost program. In our evaluation, all benchmark problems

749

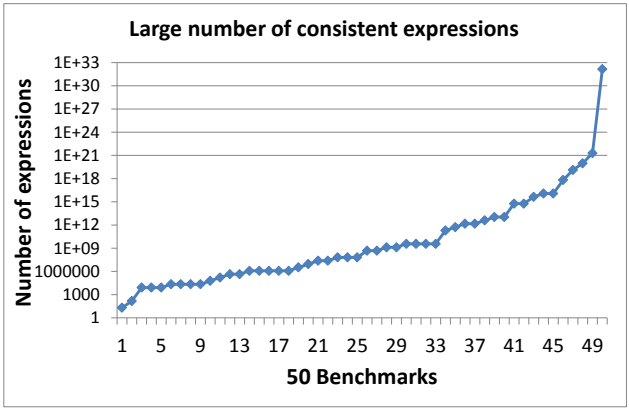
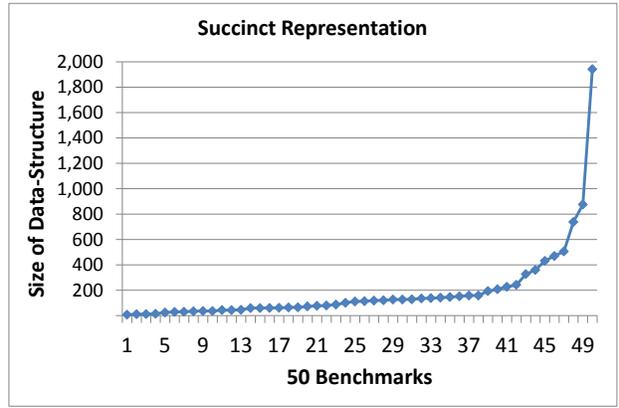

Figure 11: (a) Number of expressions consistent with given i-o examples and (b) Size of data structure to represent all consistent expressions.

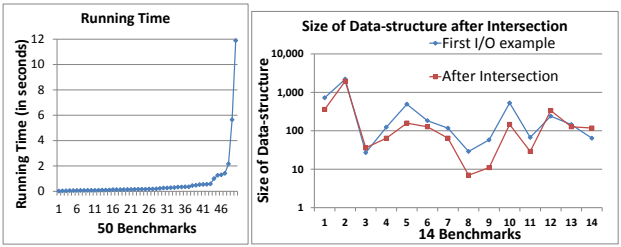

Figure 12: (a) Running time (in seconds) to learn the program and (b) Size of the data structure $D_u$ before and after performing $\texttt{Intersect}_u$.

required at most 3 examples to learn the transformation: 35 benchmarks required 1 example, 13 benchmarks required 2 examples and 2 benchmarks required 3 examples.

**Performance:** We present the running time of our synthesis algorithm to learn the desired transformation for each benchmark problem in Figure 12(a) (sorted in increasing order). Note that 88% of benchmark problems finished in less than 1 second each and 96% of problems finished in less than 2 seconds each. The experiments were performed on a machine with Intel Core i7 1.87GHz CPU and 4GB RAM.

**Size of data structure after Intersection:** Finally, we show empirically that the $\texttt{Intersect}_u$ procedure does not cause a quadratic blowup in the size of the data structure for any of our benchmark problems. We compare the sizes of the data structure corresponding to the first input-output example and the data structure obtained after performing the $\texttt{Intersect}_u$ procedure in Figure 12(b). As we can see the size of the data structure mostly decreases after intersection and increases slightly in a few cases, but it is very far from a quadratic increase in its size.

## 8. RELATED WORK

Within the database literature, our work is most closely related to the problems of *record matching*, *learning schema matches* and *query synthesis*. We have detailed some differences below, but the most significant difference is that we put these concepts together.

**Record Matching:** The task of syntactic manipulation performed before a lookup operation in our extended transformation language can be likened to the problem of *record matching*. Most of the prior work in this area [5, 12] has focused on designing appropriate similarity functions such as edit distance, jaccard similarity, cosine similarity, and HMM25. A basic limitation of most of them is that they have limited customizability. Arasu et.al. have proposed a customizable similarity measure that can either be user-programmed [1] or can be inferred from examples of matching textual records [2]. In both these cases, the underlying transformation rules only involve constant strings, e.g., US → United States. Our record matching is also inferred from examples, but it involves generalized transformation rules consisting of syntactic operations such as regular expression matching, substring, and concatenate.

**Learning Complex Schema Matches:** The problem of synthesizing semantic string manipulations is also related to the problem of finding complex semantic matches between the data stored in disparate sources. The iMAP system [4] finds the schema matches that involve concatenation of column strings across different tables using a domain-oriented approach. Another approach by Warren and Tompa [19] learns the relationships that involve concatenation of column substrings, but within a single table using a greedy approach. Our language-theoretic approach learns relationships that involve concatenation of column substrings across multiple database tables without using any domain knowledge about the column entries.

**Query Synthesis by Example:** The *view synthesis* [3, 18] problem aims to find the most succinct and accurate query for a given database view instance. The high-level goal of this work is similar to that of our inductive synthesis algorithm for the lookup transformation language $L_t$, but there are some key differences: (i) View synthesis techniques infer a relation from a large representative example view, while we infer a transformation from a set of few example rows (which is a critical usability aspect for end-users). (ii) View synthesis techniques infer the most likely relation, while our lookup synthesis algorithm infers a succinct representation of all possible hypotheses, which enables its extension to a synthesis algorithm for the language $L_u$. (iii) The technique in [3] does not consider *join* or *projection* operations.

**Text-editing Systems using Demonstrations and Examples:** QuickCode [8] is a programming by example system for automating syntactic string transformations in spreadsheets. It synthesizes programs with restricted form of regular expressions, conditionals, and loops for performing syntactic string transformations. Our work leverages QuickCode to perform semantic string transformations.

Programming-by-demonstration systems for text-editing like SMARTedit [14] or simultaneous editing [15] require the user to provide a complete demonstration or trace, where the demonstration consists of a sequence of the editor state after each primitive action, really spelling out how to do the transformation, but on a given example. This is the main reason why such systems have not yet been very successful [13]. Our system is based on Programming by Example (as opposed to Programming by Demonstration) – it requires the user to provide only the final state (as opposed to also providing the intermediate states).

The work in [10] describes a programming by example technology for learning layout transformations on tables. In contrast, this paper describes a learning algorithm for synthesizing string transformations based on table lookups.

The systems described above are structured along the general formalism described in §3; however, this paper presents an instantiation to a novel domain of semantic string manipulation based on a novel learning algorithm. None of the examples that we describe in this paper can be addressed by any of these systems (except Example 4, which can be handled by QuickCode) because they don't implement any reasoning about semantic data types.

## 9. CONCLUSION AND FUTURE WORK

Program synthesis is the task of synthesizing a program in some underlying language from specifications that can range from logical declarative specifications to examples or demonstrations. This topic has been studied extensively in AI and PL communities with the goal of easing the burden of algorithm designers or software developers. (See [7] for a recent survey.) Since program synthesis is a hard combinatorial problem, and these users write sophisticated programs, we have not yet been able to design robust tools that can provide significant value to this class of users on a daily basis. As a result, existing program synthesis techniques have not yet found significant adoption in real world. In contrast, this paper targets end-users, whose needs are much simpler compared to those of software developers. This paper presents one such tool that is ready to be deployed for use by end-users in real world. More significantly, the impact potential is huge: 500 million spreadsheet users who struggle with spreadsheets on a daily basis! We believe that further research in this area of program synthesis for end-users can potentially bring a computing revolution by democratizing the ability to effectively use computational devices.

In this paper, we have considered spreadsheet tables, which are typically small in size and lead to real-time performance of our learning algorithm. It would be interesting to consider improvements to our learning algorithm to allow for efficient handling of larger database tables (where the number of reachable strings can be huge). There might also be an opportunity for designing new interaction models where users may point out the set of relevant sub-tables.

## 10. REFERENCES

[1] A. Arasu, S. Chaudhuri, and R. Kaushik. Transformation-based framework for record matching. In *ICDE*, pages 40–49, 2008.

[2] A. Arasu, S. Chaudhuri, and R. Kaushik. Learning string transformations from examples. *PVLDB*, 2(1):514–525, 2009.

[3] A. Das Sarma, A. Parameswaran, H. Garcia-Molina, and J. Widom. Synthesizing view definitions from data. In *ICDT*, pages 89–103, 2010.

[4] R. Dhamankar, Y. Lee, A. Doan, A. Y. Halevy, and P. Domingos. iMAP: Discovering complex mappings between database schemas. In *SIGMOD*, pages 383–394, 2004.

[5] A. K. Elmagarmid, P. G. Ipeirotis, and V. S. Verykios. Duplicate record detection: A survey. *IEEE Trans. Knowl. Data Eng.*, 19(1):1–16, 2007.

[6] M. Gualtieri. Deputize end-user developers to deliver business agility and reduce costs. In *Forrester Report for Application Development and Program Management Professionals*, April 2009.

[7] S. Gulwani. Dimensions in program synthesis. In *PPDP*, pages 13–24, 2010.

[8] S. Gulwani. Automating string processing in spreadsheets using input-output examples. In *POPL*, pages 317–330, 2011.

[9] S. Gulwani, W. R. Harris, and R. Singh. Spreadsheet data manipulation using examples. In *Communications of the ACM*, 2012. To Appear.

[10] W. R. Harris and S. Gulwani. Spreadsheet table transformations from examples. In *PLDI*, pages 317–328, 2011.

[11] S. Jha, S. Gulwani, S. A. Seshia, and A. Tiwari. Oracle-guided component-based program synthesis. In *ICSE*, pages 215–224, 2010.

[12] N. Koudas, S. Sarawagi, and D. Srivastava. Record linkage: similarity measures and algorithms. In *SIGMOD*, pages 802–803, 2006.

[13] T. Lau. Why programming-by-demonstration systems fail: Lessons learned for usable ai. *AI Magazine*, 30(4):65–67, 2009.

[14] T. Lau, S. Wolfman, P. Domingos, and D. Weld. Programming by demonstration using version space algebra. *Machine Learning*, 53(1-2):111–156, 2003.

[15] R. C. Miller and B. A. Myers. Interactive simultaneous editing of multiple text regions. In *USENIX Annual Technical Conference, General Track*, pages 161–174, 2001.

[16] R. Singh and S. Gulwani. Learning semantic string transformations from examples. Technical Report MSR-TR-2012-5, Microsoft Research, 2012.

[17] R. Singh and S. Gulwani. Synthesizing number transformations from input-output examples. In *CAV*, 2012. To Appear.

[18] Q. T. Tran, C.-Y. Chan, and S. Parthasarathy. Query by output. In *SIGMOD*, pages 535–548, 2009.

[19] R. H. Warren and F. W. Tompa. Multi-column substring matching for database schema translation. In *VLDB*, pages 331–342, 2006.